\documentclass[preprint]{aastex}
\shorttitle{No evidence for a dark matter in the Galactic disk}
\shortauthors{Moni Bidin et al.}

\begin{document}

\title{NO EVIDENCE FOR A DARK MATTER DISK WITHIN 4 kpc FROM THE GALACTIC PLANE}

\author{C. Moni Bidin}
\affil{Universidad de Concepci\'on, departamento de Astronom\'ia, Casilla 160-C, Concepci\'on, Chile}
\email{cmbidin@astro-udec.cl}
\author{G. Carraro\altaffilmark{1}}
\affil{European Southern Observatory, Alonso de Cordova 3107, Vitacura, Santiago, Chile}
\altaffiltext{1}{Dipartimento di Astronomia, Universit\'a di Padova, Vicolo Osservatorio 3, I-35122, Padova, Italia}
\author{R. A. M\'endez}
\affil{Universidad de Chile, Departamento de Astronom\'ia, Casilla 36-D, Santiago, Chile}
\and
\author{W. F. van Altena}
\affil{Dept. of Astronomy, Yale University, P.O. Box 208101, New Haven, CT 06520-8101, USA}

\begin{abstract}
We estimated the dynamical surface mass density ($\Sigma$) at the solar Galactocentric distance between 2 and 4 kpc
from the Galactic plane, as inferred from the observed kinematics of the thick disk. We find $\Sigma$($z$=2~kpc)=57.6$\pm$5.8
$M_\odot$ pc$^{-2}$, and it shows only a tiny increase in the $z$-range considered
by our investigation. We compared our results with the expectations for the visible mass, adopting the most recent
estimates in the literature for contributions of the Galactic stellar disk and interstellar medium, and proposed models
of the dark matter distribution. Our results match the expectation for the visible mass alone,
never differing from it by more than 0.8 $M_\odot$ pc$^{-2}$ at any $z$, and thus we find little evidence for any
dark component. We assume that the dark halo could be undetectable with our method, but the dark disk, recently
proposed as a natural expectation of the $\Lambda$CDM models, should be detected. Given the good agreement with the
visible mass alone, models including a dark disk are less likely, but within errors its existence cannot be excluded.
In any case, these results put constraints on its properties: thinner models (scale height lower than 4~kpc) reconcile
better with our results and, for any scale height, the lower-density models are preferred. We believe that successfully
predicting the stellar thick disk properties and a dark disk in agreement with our observations could be a challenging
theoretical task.
\end{abstract}


%%%%%%%%%%%%%%%%%%%%%%%%%%%%%%%%%%%%%%%%%
%%%%%%%%%%%%%%%%%%%%%%%%%%%%%%%%%%%%%%%%%

\section{INTRODUCTION}
\label{s_intro}
Today it is widely accepted that dark matter is a fundamental component of the Universe, which plays a key role in the
processes of galaxy formation and evolution. Cosmological $N$-body simulations accurately predict the evolution of the dark
component and its actual spatial distribution \citep{Moore99}. In the $\Lambda$ cold dark matter ($\Lambda$CDM) cosmology,
spiral galaxies accrete smaller building blocks into both their spheroidal component and their stellar disk.
The presence of an old thick disk, very common among spiral galaxies \citep{Dalcanton02},
is often considered the product of one or more merging events \citep[e.g.][]{Abadi03}. \citet{Lake89} first proposed
that, as satellites are torn apart by tidal forces, they should deposit their dark matter into a flat dark structure. This
idea was recently explored by \citet[][hereafter Re08]{Read08}, \defcitealias{Read08}{Re08} who showed that the presence of a
dark disk is a natural expectation of the $\Lambda$CDM model. As a result of their simulations, they proposed that a galaxy
such as the Milky Way should host a relatively thin dark feature (exponential scale height 2.1-2.4~kpc), with a
local density at the solar position ($\rho_d$) 0.25-1.0 times that of the dark halo ($\rho_h$).
The proposed dark component is much more flattened than the dark halo, but it is still noticeably thicker than any
visible disk, because the scale height of the Galactic old Thick stellar disk is $\sim$0.9 kpc,
while younger stars and interstellar medium (ISM) form even thinner structures \citep[0.3 and 0.1 kpc, respectively,][]{Juric09}.

In the last year it became progressively accepted that, if
the $\Lambda$CDM cosmology is the correct model, dark disks should be ubiquitous in spiral galaxies.
More recently, \citet[][hereafter Pu09]{Purcell09}, \defcitealias{Purcell09}{Pu09} elaborated new models, proposing a thicker
(scale height 4.6~kpc) but less dense dark disk in the Milky Way ($\rho_d$/$\rho_h \leq$0.30).

The presence of the dark disk is strongly related to the formation of the stellar thick disk. However,
its merging origin is currently under debate, because of the difficulties of the models in reproducing all its
properties. For example, \citet{Bournaud09} argued that thick disk formation through turbulent and clumpy phases at
high redshift can explain its lack of flaring \citep{Momany06}, at variance with the merger scenario. This model would not
necessarily require the presence of a dark disk. On the other hand, the presence of a phantom disk is also an expectation
of MOND theory \citep{Milgrom83}, where the departure of gravitation from Newtonian law should
cause the detection of an additional amount of disk matter \citep{Milgrom01}. It is thus clear that the Milky Way dark disk
has become a benchmark for many theories, from gravitational law to cosmological galaxy formation, and thick disk origin.

We are performing an extensive survey to study the kinematical and chemical vertical structure of the Galactic thick disk
\citep{Carraro05}. Preliminary results were presented by \citet{Moni09a}. In this Letter we analyze the vertical trend of the
surface mass density as inferred by thick disk kinematics, in search of evidence for any dark component. The detailed analysis
of the kinematical results will be published in a later paper (C. Moni Bidin et al. 2011a, in preparation, hereafter Paper~I),
and the full study of the hypothesis and equations presented here will follow (C. Moni Bidin et al. 2011b, in preparation,
hereafter Paper~II).

%%%%%%%%%%%%%%%%%%%%%%%%%%%%%%%%%%%%%%%%%
%%%%%%%%%%%%%%%%%%%%%%%%%%%%%%%%%%%%%%%%%

\section{OBSERVATIONAL DATA}
\label{s_data}
Our investigation is based on a sample of $\sim$1200 red giants defined by \citet{Girard06}, vertically distributed
with respect to the Galactic plane in a cone of 15 degrees radius centered on the south Galactic pole. All
objects have Two Micron All Sky Survey (2MASS) photometry \citep{Skrutskie06} and absolute proper motions from the SPM3
catalog \citep{Girard04}.
The sample was selected by applying a color cut in the infrared color-magnitude diagram, to isolate intermediate-metallicity
thick disk stars. Main-sequence (MS) dwarfs were excluded both by a sloped cut at fainter magnitudes,
which excludes all but the nearest ($d\leq$63~pc) dwarfs, and by conservative kinematical criteria imposing a stellar
velocity lower than the local escape velocity \citep[550 km s$^{-1}$; see][for more details]{Girard06}.
We collected high-resolution Echelle spectra for two-thirds of the Girard et
al.'s sample, during 38 nights at La Silla and Las Campanas observatories. The distribution
of proper motions and colors of this sub-sample was analyzed, to ensure that no selection effect was introduced. We
visually inspected all the spectra and excluded the residual low-metallicity stars ([Fe/H]$\leq -$1.5), most probably
halo contaminators, and misclassified
dwarfs. Radial velocities (RVs) were measured for all the stars with a typical error of 0.5-0.7 km s$^{-1}$,
cross-correlating each spectrum with standard stars
observed during the same runs \citep{Tonry79}. Distances were estimated with a color-absolute magnitude relation
calibrated on the red giant branch of 47\,Tuc, whose stellar population is very similar to the
intermediate-metallicity Galactic thick disk \citep{Wyse05}. Finally, the three velocity components in the Galactic
cylindrical coordinate system ($U,V,W$), and their associated errors, were calculated for each star from its proper motion,
RV, and distance, and the uncertainties on these quantities.
The error in distance was $\sim 20$\% \citep{Moni09b}, while the error in proper motions was fixed to 3 mas yr$^{-1}$
\citep[T. M. Girard, 2009, private communication, see also][for a detailed discussion]{Girard06}.
The mean errors in (U,V,W) thus increased from $\sim$(7,36,38) km s$^{-1}$ at $z$=2 kpc, to $\sim$(10,61,66) km s$^{-1}$
at 4 kpc. The details of observations, data reduction, and RV measurements were presented by \citet{Moni09b},
and they will be fully discussed in Paper~I.

\begin{figure}
\epsscale{1.}
\plotone{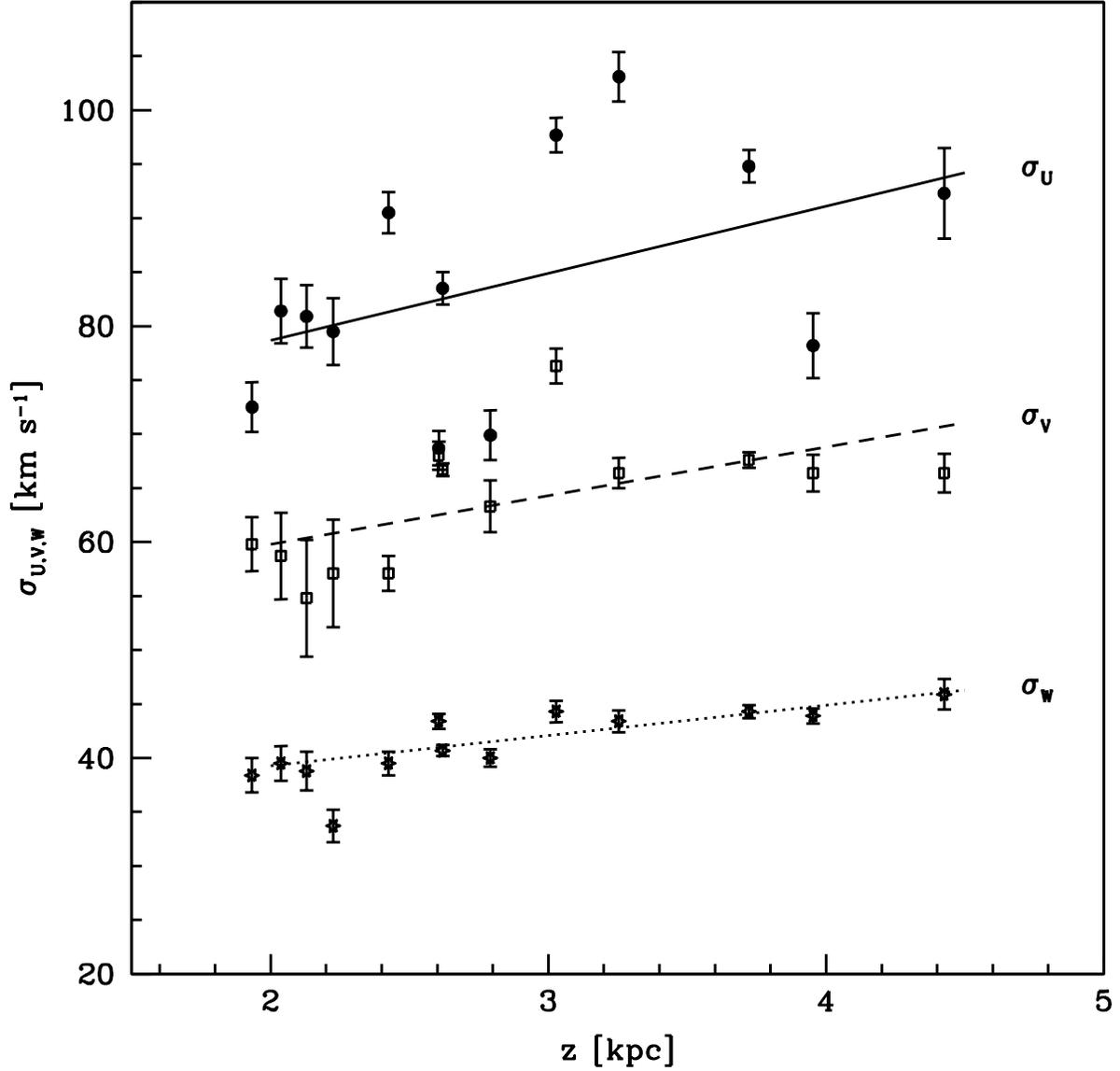}
\caption{Dispersions of the three velocity components, as a function of Galactic height, and their least-squares
fit. Full dots and line: $\sigma_U$; open squares and dashed line: $\sigma_V$;
open stars and dotted line: $\sigma_W$.
The errors in $z$ ($\leq$100 pc at 4~kpc), given by the statistical error on the mean, are omitted for clarity.
\label{f_dispersions}}
\end{figure}

\begin{figure}
\epsscale{1.}
\plotone{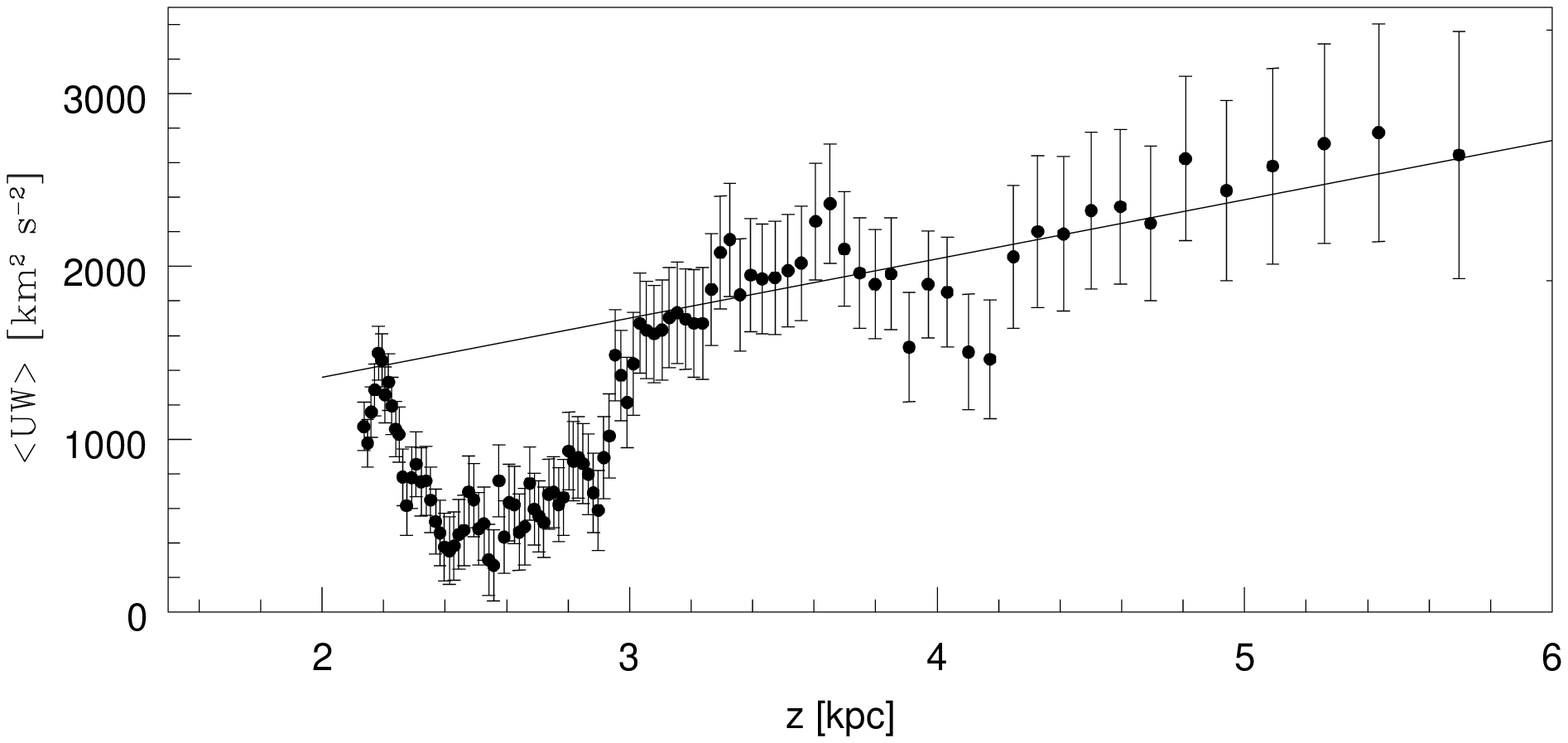}
\caption{Measured cross-term $\overline{UW}$ as a function of distance from Galactic plane. The line indicates the least
squares fit obtained omitting the depression at 2.5-3~kpc. \label{f_uw}}
\end{figure}

In this Letter, we restrict our analysis to the $\sim$300 stars beyond 2~kpc from the Galactic plane, to avoid
contamination by the thin disk. We considered only stars with $\vert W\vert\leq$150, $\vert U\vert\leq$300,
and $-500\leq V\leq$300 km s$^{-1}$, to exclude halo stars and/or bad measurements. We did not apply a cut in velocity
errors, because it systematically excludes high-velocity stars, biasing the results: the propagation of the uncertainty in
distance in fact introduces a term proportional to the velocities themselves. The sample was binned with respect to distance from
the Galactic plane $z$, following three different criteria: three large bins of 85 stars each, five bins of 50, and five
of 45 stars. The dispersion in the three velocity components was calculated in each bin, fitting the corresponding probability
plot with a linear relation (\citealt{Lutz92}; see also \citealt{Bochanski07}, for an application to a very similar astrophysical
case).
In the nearest bins ($z \leq$2.3~kpc) the fit was performed outside the $\pm1\sigma$ range, to avoid the residual thin disk
contamination (5\%-10\% according to our estimates). At increasing distance an overestimate was a serious possibility because
of some outliers, most probably stars with an incorrect distance and/or proper motion. We consequently excluded from the
fit the points on the wings of the distribution showing clear deviation from linearity, i.e.
outliers departing from the underlying normal distribution. The mean error in each bin was quadratically subtracted
to derive the final estimate of the dispersion. Finally, the dispersion in each velocity component was
plotted against the average $z$ of the bin, and a least-squares linear relation was fitted to derive the vertical trend
of dispersions.

The results were very similar in the three binning criteria: differences on the derived quantities were smaller than
1.5$\sigma$, and in most cases agreed within 1$\sigma$. We therefore gathered the
information of all the bins together, fitting a final plot comprising 13 points. The results are shown in
Figure~\ref{f_dispersions}. We derive a small vertical gradient of $\sigma_U$, $\sigma_V$, and
$\sigma_W$ of 6.2$\pm$3.7, 4.5$\pm$1.9, and 2.8$\pm$0.5 km s$^{-1}$ kpc$^{-1}$, respectively.  The statistical
uncertainty of the least square solution was assumed as the error on the linear profile parameters. These could be
underestimated because of the correlation of the fitted points, but they are only about 30\% higher when
fitting the uncorrelated bins of 50 stars, and this has only minimal impact on the final results,
because the uncertainties on the thick disk parameters dominate the error budget.

Around $z$=2.5~kpc we find a sudden deviation from linearity of the cross-term of the dispersion matrix, $\overline{UW}$. This
feature could be due to a stellar sub-structure, such as a comoving group of stars. The use of only 13 bins is inadequate
to reveal the general trend, hence we binned the data in overlapping groups of 50 stars at steps of 2 stars, as shown in
Figure~\ref{f_uw}. We finally derived the linear relation required for our calculations excluding the ''depression" at 2.5~kpc.
The line is a good fit outside this feature.

%%%%%%%%%%%%%%%%%%%%%%%%%%%%%%%%%%%%%%%%%
%%%%%%%%%%%%%%%%%%%%%%%%%%%%%%%%%%%%%%%%%

\section{THE THEORY}
\label{s_theory}

\renewcommand{\theenumi}{\Roman{enumi}}
\renewcommand{\labelenumi}{\theenumi}

Our formulation is based on the following assumptions:
\begin{enumerate}
\item {\it Steady state}. The thick disk is in equilibrium with the Galactic potential, as expected for an old stellar
population. Therefore, all temporal derivatives are set to zero. \label{hypo_steady}
\item {\it Locally flat rotation curve}. The rotation curve is assumed flat at the solar Galactocentric distance.
\item {\it No net radial or vertical stellar flux}. The mean radial and vertical velocity components are zero, while the
rotational component shows a non-null lag \citep{Chiba00,Girard06}.
\item {\it Exponential radial dispersion profiles}. All the velocity dispersions decrease with $R$ following an exponential
law, with a scale length h$_{R,\sigma}$ equal to the one observed for the mass density (h$_{R,\rho}$).
\label{hypo_expo}
\item {\it Vertical constancy of scale lengths}. h$_{R,\rho}$ and h$_{R,\sigma}$ do not depend on $z$ at
the solar position. This is observationally verified for the mass density \citep{Cabrera05}, and it is assumed
valid for the velocity dispersions, because their radial behavior is linked to the mass distribution by assumption
(\ref{hypo_expo}).
\item {\it Null cross term on the Galactic plane}. $\overline{UW}(z=0)$=0. This hypothesis is required for symmetry reasons
\citep[see e.g.][]{Bienayme09b}, and it is observationally confirmed by \citet{Fuchs09}. \label{hypo_uw}
\end{enumerate}
The hypothesis (\ref{hypo_expo}) is observationally confirmed for $\sigma_W$ \citep{Kruit82}. Its extension to
the other components relies on the controversial assumption of constant anisotropy, i.e.
$\frac{\partial}{\partial R} \frac{\sigma_U}{\sigma_W}=0$. We have no information about the radial trend of the
velocity dispersions, but this assumption is supported by observations \citep{Lewis89}, and numerical simulations for
$R\leq$9~kpc \citep{Cuddeford92}.

Integrating the Poisson equation from $-z$ to $z$, assuming that the vertical component of the
force F$_z$ is null on the plane, we obtain:
\begin{equation}
-2\pi G\Sigma(z)=\int_{0}^{z}\frac{1}{R}\frac{\partial}{\partial R}(R F_R)dz+F_z(z),
\label{eq_Pois}
\end{equation}
where $\Sigma (z)$ is the surface mass density between $\pm z$, and F$_R$ is the radial component of the force.
Inserting the Jeans equations in Equation~(\ref{eq_Pois}),
making use of our hypotheses (\ref{hypo_steady})-(\ref{hypo_uw}), and calculating
simple derivatives and integrals, we eventually obtain the final expression:
\begin{equation}
\Sigma(z)=\frac{1}{2\pi G}\Bigl{[}
k_{1}\cdot\int_{0}^{z}\sigma_{U}^{2}dz+k_{2}\cdot\int_{0}^{z}\sigma_{V}^{2}dz+k_{3}\cdot\int_{0}^{z}\overline{UW}dz+
k_{4}\cdot \overline{UW}+\frac{\sigma^{2}_{W}}{h_{z,\rho}}-\frac{\partial \sigma^{2}_{W}}{\partial z}\Bigl{]}
\label{eq_final},
\end{equation}
where h$_{z,\rho}$ is the thick disk exponential scale height, and:
\begin{eqnarray}
&&k_{1}=\frac{3}{R_{\odot}\cdot h_{R,\rho}}-\frac{2}{h_{R,\rho}^{2}}, \\
&&k_{2}=-\frac{1}{R_{\odot}\cdot h_{R,\rho}}, \\
&&k_{3}=-\frac{1}{h_{z,\rho}}\cdot \Bigl{(}\frac{1}{R_{\odot}}-\frac{1}{h_{R,\rho}}\Bigl{)}, \\
&&k_{4}=\frac{3}{h_{R,\rho}}-\frac{2}{R_{\odot}}.
\end{eqnarray}
To calculate the surface density we must insert into Equation~(\ref{eq_final}) the vertical trends of the
kinematical quantities $\sigma_U$, $\sigma_V$, $\sigma_W$, and $\overline{UW}$,
plus the three parameters R$_{\odot}$, h$_{z,\rho}$, and h$_{R,\rho}$.
We fixed R$_{\odot}$=8.0$\pm$0.3~kpc, and we defined the thick disk scale height and length as the mean of
about 20 literature estimates \citep[see][and Paper~II for the bibliographical references]{Moni09b},
obtaining h$_{R,\rho}$=3.8$\pm$0.2~kpc and h$_{z,\rho}$=0.90$\pm$0.08~kpc. The quoted errors
are given by the error on the mean, that should be the most appropriate statistic when averaging many uncorrelated
measurements. However, they could be too small because of the traditional uncertainty on these parameters,
although the estimates converged considerably in the last years. We therefore also considered an uncertainty of 0.4 and
0.12 kpc, respectively, comprising all the measurements of the last decade within $\pm 2\sigma$ ($\sim$70\% within
$\pm 1\sigma$). These larger errors necessarily decrease the significance of the results, but without altering the general
conclusions, as can be deduced from Figure~\ref{f_darkdisk}.
The error on $\Sigma (z)$ was calculated from the propagation of errors of all the quantities in Equation~(\ref{eq_final}).

In deriving Equation~(\ref{eq_final}), we assumed the kinematical quantities as symmetric with respect to the
plane, so that their integrals between $-z$ and $z$ are twice the product of integration between 0
and $z$. This is easily justified for the dispersions, but it may not be the case for $\overline{UW}$.
We therefore also considered an antisymmetric cross-term, where the third term of
Equation~(\ref{eq_final}) vanishes. As shown in Figure~\ref{f_results}, a symmetric cross-term fails to return physically
meaningful results, because it violates two minimum requirements: the surface density must at least account for the known
visible matter, and it cannot decrease with $z$. We will therefore assume the cross-term as antisymmetric, and consider
only the results obtained under this hypothesis hereafter.

%%%%%%%%%%%%%%%%%%%%%%%%%%%%%%%%%%%%%%%%%
%%%%%%%%%%%%%%%%%%%%%%%%%%%%%%%%%%%%%%%%%

\section{RESULTS}
\label{s_results}

\begin{figure}
\epsscale{1.}
\plotone{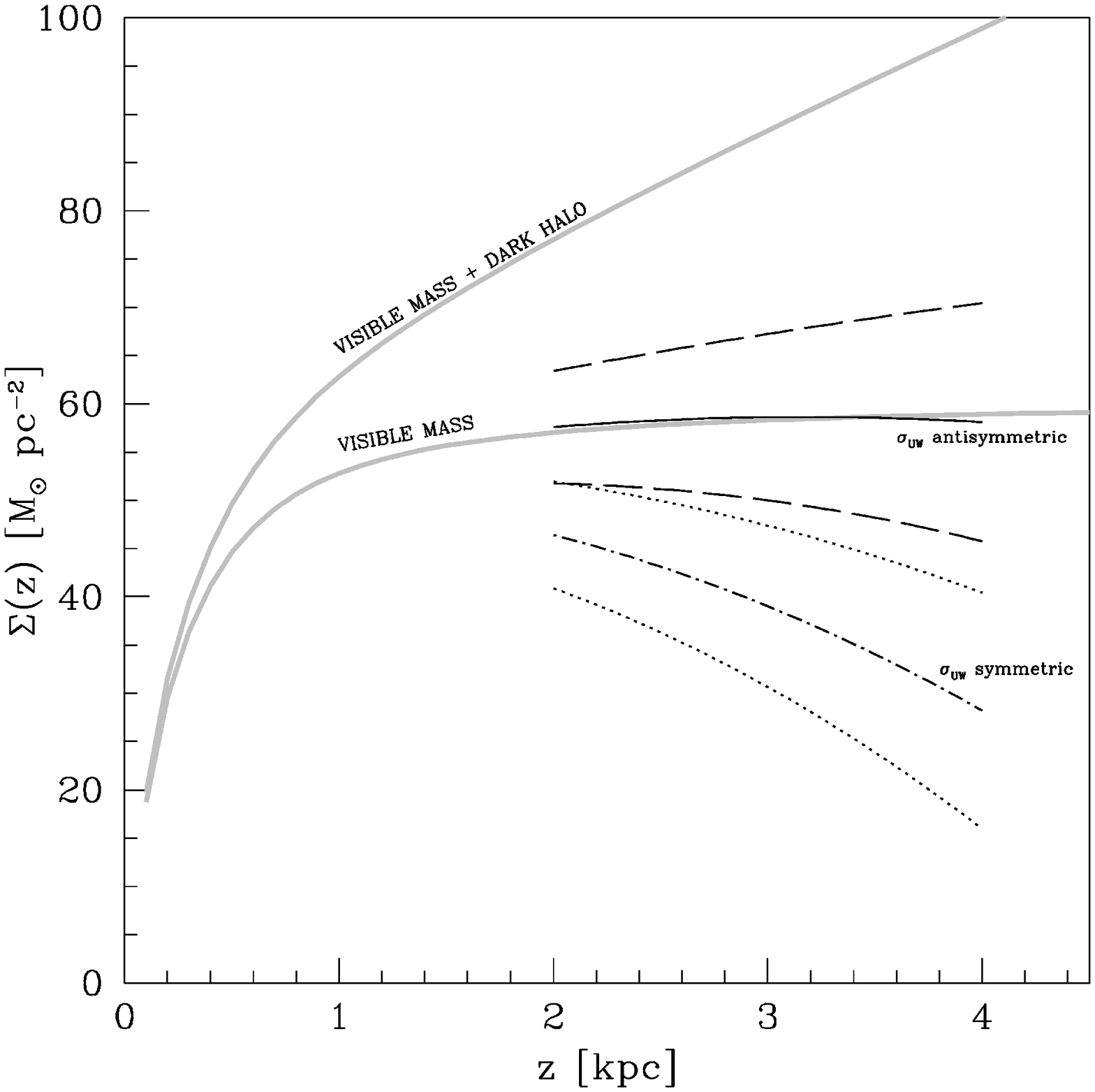}
\caption{The surface density vertical profile derived from our observations. The full line shows the results under the
assumption of cross-term antisymmetry, while the dash-dotted line indicates the results for a symmetric cross-term.
The dashed and dotted lines indicate the corresponding 1$\sigma$ strips. Light gray curves reprensents the
expectations of the visible mass alone (lower curve), and of a visible mass+dark halo model (upper curve).
\label{f_results}}
\end{figure}

\begin{figure}
\epsscale{1.}
\plotone{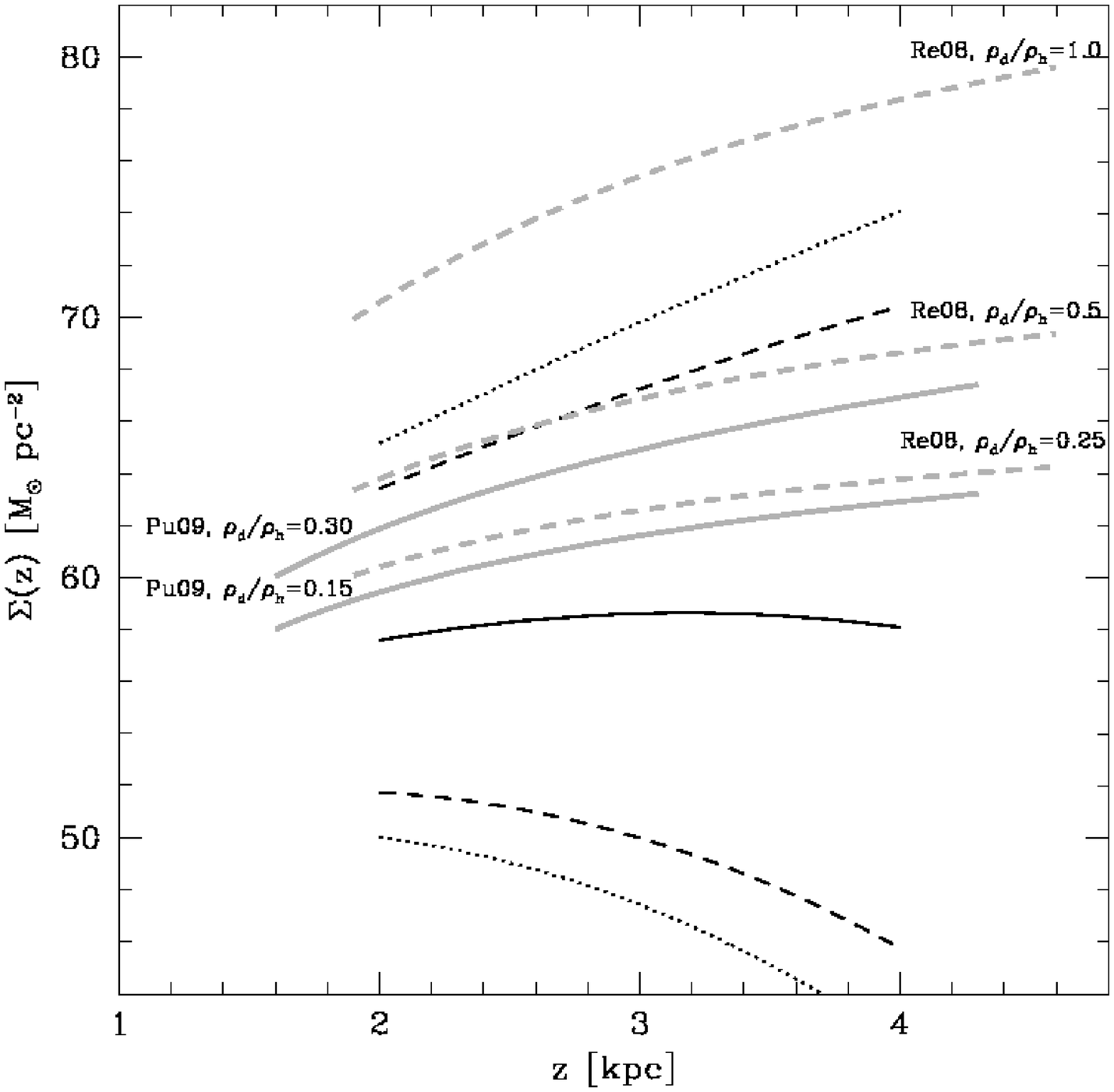}
\caption{Comparison of our results with the expectations of dark disk models. The black full line shows the observational result
with its 1$\sigma$ errors ({\bf dashed} lines) as in Figure~\ref{f_results}. {\bf The dotted lines indicate the 1$\sigma$ strip
when the enhanced errors on the thick disk parameters, discussed in the text, are considered}. The light grey curves are models
where a dark disk is added to the visible mass, basing on the models proposed by \citetalias{Read08} (dashed lines,
h$_{z,D}$=2.4~kpc and three different values of local density), and by \citetalias{Purcell09} (full line,
h$_{z,D}$=4.6~kpc, two values of local density).
\label{f_darkdisk}}
\end{figure}

Our results are shown in Figure~\ref{f_results}. We find $\Sigma$(2~kpc)=57.6$\pm$5.8 $M_\odot$ pc$^{-2}$, and the
curve is nearly flat in the whole range, while the error increases constantly to $\sim$12 $M_\odot$ pc$^{-2}$ at 4~kpc.
The solution increases 1~$M_\odot$ pc$^{-2}$ between 2 and 3.2~kpc, then it turns slightly downwards. However, the decrease
is so small (0.55 $M_\odot$ pc$^{-2}$) that a small refinement of any parameter
would correct it, for example increasing h$_{R,\rho}$ by 0.1~kpc. Moreover, this problem is present only in the
lower half of the family of solutions defined by the $\pm 1\sigma$ strip. We did not alter the input parameters to amend it,
because this would add a high degree of arbitrariness to the results.

In Figure~\ref{f_results} we overplot a model of surface mass density due to the visible mass.
The thin and thick disk exponential scale heights, the halo shape, and the local thick disk and halo normalization
were taken from \citet{Juric09}. We included a thin layer (100 pc) of ISM 
contributing 13 $M_\odot$ pc$^{-2}$ \citep{Holmberg00}, and the thin disk density on the plane was normalized by the
requirement $\Sigma_\mathrm{disk}$(1.1~kpc)=40 $M_\odot$ pc$^{-2}$ \citep{Holmberg04}. This quantity includes all disk stellar
components and remnants, and it is currently the best estimate often assumed in Galactic mass models
\citep{Dehnen98,Olling01,Weber09}.

The agreement between the visible mass and our dynamical solution is striking, and there is no need to invoke
any dark component. In Figure~\ref{f_results} we also plot a model including a dark halo
\citep{Olling01}. The curve is completely incompatible with our results, both for its high value and steep slope.
One can argue that we are actually deducing the mass from the change in the gravitation potential: the dark
halo could be too uniform and extended to cause a detectable change on stellar kinematics in a range of 4~kpc. This point
requires a detailed analysis that is beyond the scope of this Letter, and we
leave it to a later study (Paper~II). Here we assume that the dark halo is undetectable and focus on the dark disk, for
which this explanation is not viable.

\begin{figure}
\epsscale{1.}
\plotone{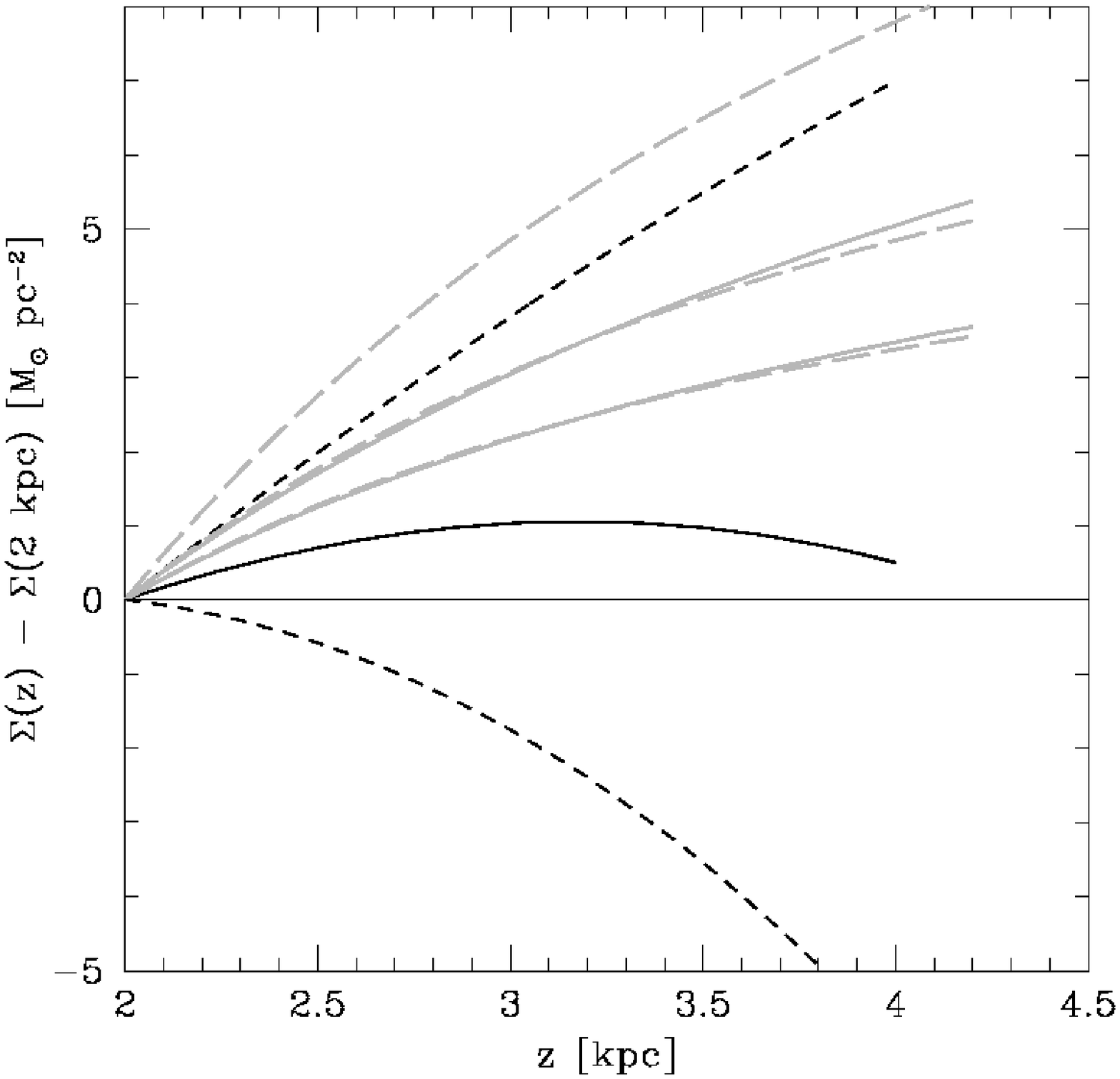}
\caption{Same as Figure~\ref{f_darkdisk}, but for the increment of the surface mass density as a function of $z$, with
respect to 2~kpc, instead of its absolute value. \label{f_increment}}
\end{figure}

Figure~\ref{f_darkdisk} compares our results with models where a dark disk is considered. They require the
definition of two additional parameters: the exponential scale height of the dark component h$_{z,D}$, and its
density at $z$=0, normalized with respect to the dark halo, $\rho_d$/$\rho_h$. We considered two families of
models, a relatively thinner (h$_{z,D}$=2.4~kpc) dark disk with $\rho_d$/$\rho_h$=0.25-1.0, as
proposed by Re8, and a thicker, less dense one (h$_{z,D}$=4.6~kpc, $\rho_d$/$\rho_h$=0.15-0.30), from
\citetalias{Purcell09}.
The measured surface density matches the baryonic mass alone, and any curve including the dark disk departs from the
central solution, being therefore less likely. Hence, there is no evidence for a flattened dark component, but its expected
contribution is small compared
to the errors and its existence cannot be completely ruled out. However, not all combinations of dark disk parameters
are permitted by our observations: the \citetalias{Purcell09}
solutions fall in the 1$\sigma$-strip, but only the less dense models of \citetalias{Read08} do. This is entirely due
to the lower densities of the first family of models. In fact, for a
fixed $\rho_d$/$\rho_h$, the thick \citetalias{Purcell09} models depart more from the observed curve, and
they require a higher increment of $\Sigma(z)$ compared to \citetalias{Read08} models (see Figure~\ref{f_increment}).
We can thus derive important constraints: a lower density is favored in all cases, and thinner disks should
be preferred, while thick and dense dark disks (h$_{z,D}\geq$4~kpc, $\rho_d$/$\rho_h\geq$0.5) are
less likely.
Numerical simulations showed that low-latitude merging of massive satellites are required to form a heated disk
kinematically similar to the Galactic thick disk \citep{Villalobos09}, but these events produce denser dark disks
\citepalias{Purcell09}. In addition, \citetalias{Read08} showed that the observed density of the stellar thick disk can be
reproduced only by a few models, all implying a more massive dark disk ($\rho_d$/$\rho_h\geq$0.4).
In summary, correctly predicting the thick disk kinematical {\em and} dark disk properties is a challenging task.
Models for thick disk formation alternative to the merging scenario \citep[e.g.,][]{Bournaud09} are preferred, while the
MONDian prediction of \citet{Bienayme09a}, i.e. a 60\% increase of disk mass due to a phantom disk
($\sim$30 $M_\odot$ pc$^{-2}$ at 2~kpc) is in contradiction with our results.

It could be argued that our model of visible mass, it relies on poorly-constrained quantities, such as the ISM contribution and
$\Sigma_\mathrm{disk}$(1.1~kpc), and a downward correction of these parameters would shift all the model curves to lower values.
Therefore, in Figure~\ref{f_increment} we analyzed both the expected and measured increment
of $\Sigma (z)$, i.e. the surface density of the mass comprised between $z$ and 2~kpc. This quantity
is completely independent of the ISM at these Galactic heights, and the assumed value of $\Sigma_\mathrm{disk}$(1.1~kpc)
introduces a negligible uncertainty ($\leq$0.15 $M_\odot$ pc$^{-2}$). Figure~\ref{f_increment} shows exactly the same
situation analyzed previously, and our conclusions should be considered independent of the details and uncertainties of the
modeled visible mass.

%%%%%%%%%%%%%%%%%%%%%%%%%%%%%%%%%%%%%%%%%
%%%%%%%%%%%%%%%%%%%%%%%%%%%%%%%%%%%%%%%%%

\section{CONCLUSIONS}
\label{s_conclusions}

We estimated the Galactic surface mass density between 2 and 4~kpc from the plane, finding
$\Sigma$(2~kpc)=58$\pm$6 $M_\odot$ pc$^{-2}$, and a nearly flat curve. Our results strikingly match the visible mass alone,
and we do not detect evidence for any dark component, although the dark halo could have passed unseen.
There is no compelling evidence for a dark disk, but within the errors of our
investigation, its existence cannot be completely excluded. We derive important constraints on its expected properties: lower
densities ($\rho_d$/$\rho_h\leq$0.25) should be preferred in any case, and a thin dark disk
(h$_{z,D}\leq$2.5~kpc) better reconciles with observations. A thick and dense dark disk (h$_{z,D} \geq$4~kpc,
$\rho_d$/$\rho_h\geq$0.5) should be excluded. Any merging model aiming to reproduce the formation of the
Galactic thick disk and a flat dark component will need to consider the constraints from our investigation.

\acknowledgments
The authors are grateful to V. Korchagin for his assistance.
CMB acknowledges the Chilean Centro de Excelencia en Astrof\'isica y Tecnolog\'ias Afines (CATA), D. Geisler for his
reading of the manuscript, and F. Mauro for useful discussions. RAM  acknowledges support by the Chilean Centro
de Astrofisica FONDAP (No. 15010003), FONDECYT (No. 1070312), and by CATA (PFB-06).  All authors acknowledge partial support from
the Yale University/Universidad de Chile collaboration.  The SPM3 catalog was funded in part by grants from the US National
Science Foundation, Yale University and the Universidad Nacional de San Juan, Argentina.

{\it Facilities:} \facility{Du Pont (ECHELLE)}, \facility{Magellan:Clay (MIKE)},
\facility{Euler1.2m (CORALIE)}, \facility{Max Plank:2.2m (FEROS)}

\end{document}